\documentclass[10pt,twocolumn,a4paper]{esaAI}

\usepackage{aas_macro}
\usepackage{xcolor}
\usepackage{url}
\usepackage[USenglish,british,american,australian,english]{babel}

\title{A machine learning approach for computing solar flare locations in X-rays on-board Solar Orbiter/STIX}

\def\authorEmail{paolo.massa@fhnw.ch}

\author[1]{Paolo Massa\thanks{Corresponding author. E-Mail: \authorEmail}}
\author[1,2]{Simon Felix}
\author[1,2]{László István Etesi}
\author[3]{Ewan C. M. Dickson}
\author[1]{Hualin Xiao}
\author[1,4]{Francesco P. Ramunno}
\author[1]{Merve Selcuk-Simsek}
\author[1]{Brandon Panos}
\author[1]{Andr\'e Csillaghy}
\author[1,5]{Säm Krucker}
\affil[1]{Institute for Data Science, University of Applied Sciences and Arts Northwestern Switzerland, Bahnhofstrasse 6, 5210 Windisch, Switzerland}
\affil[2]{Ateleris GmbH, Neumarkt 1, 5200 Brugg, Switzerland}
\affil[3]{Institute of Physics, University of Graz, A-8010 Graz, Austria}
\affil[4]{Department of Computer Science, University of Geneva, 1211 Geneva, Switzerland}
\affil[5]{Space Sciences Laboratory, University of California, 7 Gauss Way, 94720 Berkeley, USA}

\begin{document}

\makeCustomtitle

\begin{abstract}

The Spectrometer/Telescope for Imaging X-rays (STIX) on-board the ESA Solar Orbiter mission retrieves the coordinates of solar flare locations by means of a specific sub-collimator, named the Coarse Flare Locator (CFL). 
When a solar flare occurs on the Sun, the emitted X-ray radiation casts the shadow of a peculiar ``H-shaped’’ tungsten grid over the CFL X-ray detector. 
From measurements of the areas of the detector that are illuminated by the X-ray radiation, it is possible to retrieve the $(x,y)$ coordinates of the flare location on the solar disk.

In this paper, we train a neural network on a dataset of real CFL observations to estimate the coordinates of solar flare locations.
Further, we apply a post-training quantization technique specifically tailored to the adopted model architecture. 
This technique allows all computations to be in integer arithmetic at inference time, making the model compatible with the STIX computational requirements.
We show that our model outperforms the currently adopted algorithm for estimating the flare locations from CFL data regarding prediction accuracy while requiring fewer parameters.
We finally discuss possible future applications of the proposed model on-board STIX.

\end{abstract}

\section{Introduction}
The Spectrometer/Telescope for Imaging X-rays (STIX; \cite{krucker2020spectrometer}) is the telescope of the ESA Solar Orbiter mission \cite{2020A&A...642A...1M} dedicated to the observation of solar flares.
STIX measures the flaring X-ray radiation between 4 and 150 keV by means of Cadmium-Telluride (CdTe) detectors \cite{2012NIMPA.695..288M}, which count the number of incident X-ray photons.
Specifically, 30 STIX detectors are mounted behind pairs of tungsten grids, which modulate the incident X-ray radiation and encode information on the morphology and location of the X-ray source(s).
The units consisting of a detector and a grid pair, named \emph{sub-collimators}, provide indirect information on the flaring X-ray source(s) in terms of a limited number of corresponding 2D Fourier components \cite{2023SoPh..298..114M}.
Therefore, once the STIX data are downloaded to the ground, it is possible to reconstruct the image of the X-ray source(s) for a specific flaring event by solving a Fourier inversion problem \cite{pianabook}.

\begin{figure*}[t]
\centering
\includegraphics[width=.87\textwidth]{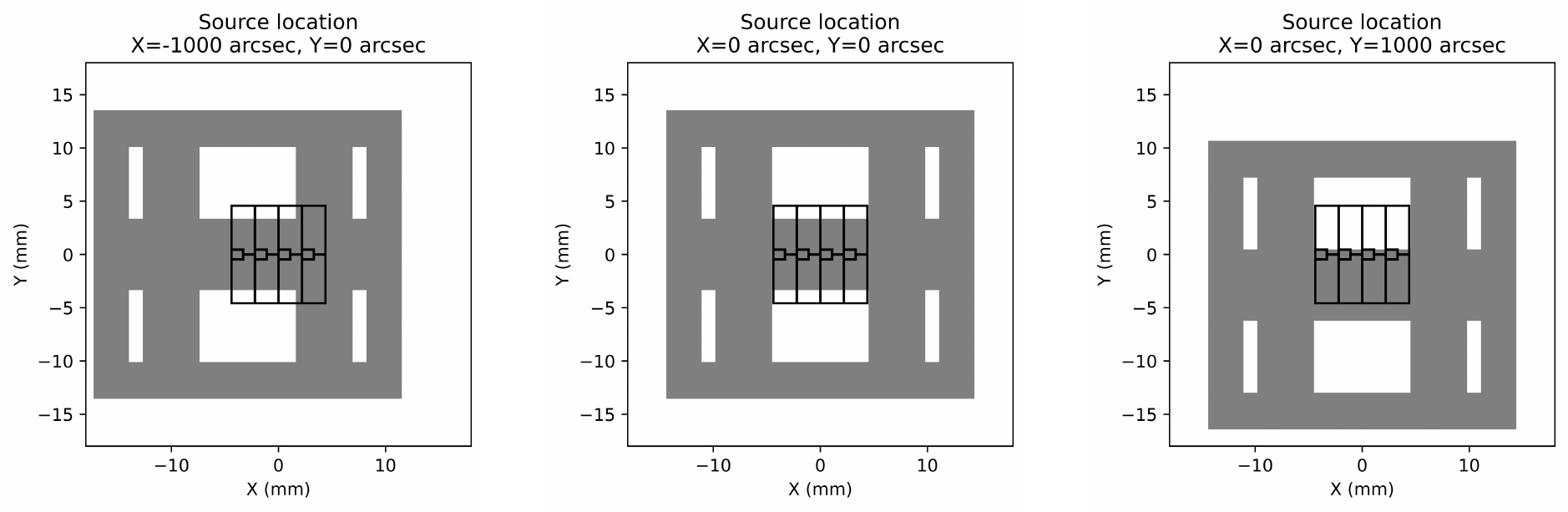}
\caption{Simulated shadow of the CFL front grid that is cast on the detector surface by three point-like flare sources located in as many different locations on the solar disk. 
The detector edges are plotted in black, while the grid shadow is overlaid in gray.
The simulated flare locations are $(x,y) = (-1000'',0'')$, $(x,y) = (0'',0'')$, and $(x,y) = (0'',1000'')$ from left to right, respectively.
The simulation is performed by means of the pystixsim library \cite{xiao_2024_10973277}.}
\label{fig:1}
\end{figure*}

STIX contains an additional sub-collimator, named the Coarse Flare Locator (CFL), which provides information on the $(x,y)$ coordinates of the flare location on the solar disk with an accuracy of a couple of arcminutes \cite{krucker2020spectrometer}.
The CFL sub-collimator consists of a front grid with a peculiar ``H-shaped'' design, while the rear grid is open.
The location of the X-ray shadow cast by the front grid on the detector surface varies depending on the flare's location on the Sun.
Therefore, the $(x,y)$ coordinates of the flare location are uniquely determined from measurements of the illuminated areas of the CFL detector (see \cref{fig:1}).

The estimates of the flare location provided by the STIX CFL are extremely important for several reasons.
First, they can be sent to other Solar Orbiter instruments for coordinated flare observations. 
Second, the flare locations estimated on-board are sent to the ground as quick-look data products with reduced latency (within less than a day) \cite{krucker2020spectrometer,2023A&A...673A.142X}.
Since, due to telemetry constraints, the STIX science data are available only after several weeks, the flare locations saved in the quick-look data products are important for preliminary analysis and for performing appropriate data requests from the ground \cite{2023A&A...673A.142X}. 
Finally, it is the first time a device like the STIX CFL operates within a space mission.  
Therefore, the flare locations computed on-board STIX are extremely valuable to demonstrate that this technique works as expected in a deep space environment and, therefore, a similar system could be used in future space weather missions.

In this work, we develop a machine learning approach to predict the coordinates of the flare location from CFL data.
We construct a dataset of real CFL measurements and label them with the corresponding location obtained from images of flares reconstructed from STIX data.
We train a multi-layer perceptron (MLP) and apply a post-training model quantization technique specifically developed for the adopted MLP architecture.
In this way, at inference time, our model applies only integer arithmetic and, therefore, satisfies the computational requirements to be run on-board STIX.
We compare the performance of our model with that of the currently used CFL algorithm, showing that the MLP outperforms the CFL algorithm in terms of prediction accuracy.

The paper is organized as follows.
\cref{sec:2} contains the details of the dataset we constructed, the MLP architecture and training, the CFL algorithm, and the comparison between MLP and the CFL algorithm performances on the test set.
Finally, \cref{sec:3} contains our discussions on the results and future work.

\section{Results}\label{sec:2}

We define an MLP with three hidden layers, each consisting of 100 neurons. 
The activation function of each neuron is a Rectified Linear Unit (ReLU \cite{bishop2006pattern}). 
Our MLP takes a 9-dimensional input. 
The first eight entries are the number of counts registered in the large pixels\footnote{A STIX detector is partitioned into 12 units, called \emph{pixels}, which count the number of incident X-ray photons independently (see \cref{fig:1}). In particular, there are 8 large pixels (top and bottom row) and 4 small ones (central row).} of the CFL detector.
The ninth entry is the sum of the counts registered by a separate imaging detector (labeled as \textit{7b}; see Table 2 in \cite{krucker2020spectrometer}), which has been chosen since it shares similar calibration properties as the CFL.
This ninth value provides an independent estimate of the total flux (i.e., total number of counts per unit of area) that should be recorded in the CFL pixels.
For example, let's consider the case when one of the two rows of the CFL pixels is fully covered by the mask shadow, as shown in the right panel in \cref{fig:1}. 
By comparing the number of counts recorded by the illuminated pixels of the CFL with the total number of counts recorded by detector \textit{7b}, the MLP can determine the amount of illuminated area of the CFL pixels and, therefore, estimate the flare location correctly.
The MLP's output layer consists of two neurons, which represent the $x$ and $y$ coordinates of the flare location on the solar disk.
No activation function is used in the output layer.

\subsection{Dataset}\label{sec:dataset}
\begin{figure}[t]

\centering
\includegraphics[width=.43\textwidth]{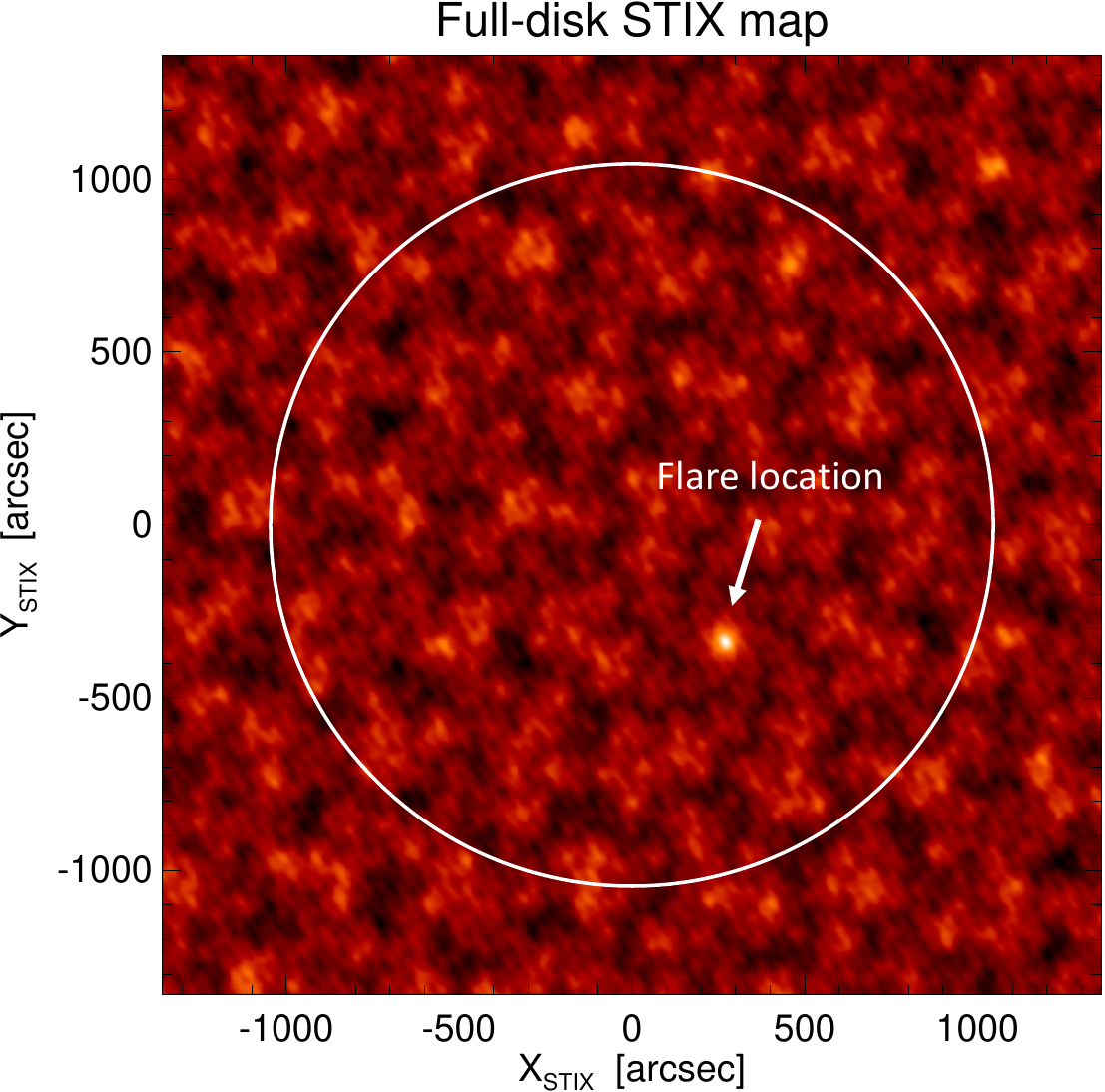}
\caption{Full-disk image reconstructed from STIX data. The solar limb is plotted as a white circle, and the flare location $(x,y) \simeq (250'',-350'')$ is indicated with a white arrow.}
\label{fig:bp}

\end{figure}

We consider 2150 solar flares recorded by STIX between 2021 March 19 and 2023 August 28. 
In this proof of concept study, we do not consider flaring events with locations outside the full-sensitivity FOV of STIX (see Section 5 in \cite{2023SoPh..298..114M}) since they are quite rare (fewer than 10\% of the total events) and make the neural network training more challenging.
For each event we consider the raw counts measured by the STIX detector pixels in 32 energy channels and in several consecutive time bins, whose number varies from event to event. 
We sum the counts recorded between 5 and 9 keV, and group the data in bins larger or equal to 8 s.
We extract the 9-dimensional vector of each bin to be used as input to the MLP.
At the same time, the flare's location, which will be used to train the MLP, is determined for each time bin from measurements provided by the STIX imaging detectors.
We recall that STIX does not provide us directly with images of the flaring X-ray radiation.
Rather, STIX measures a limited number of 2D Fourier components of the signal; therefore, an image reconstruction process is needed to obtain the flare image from STIX measurements.
As shown in Figure \ref{fig:bp}, we reconstruct a full-disk map from STIX data using a direct Fourier inversion method, named Back Projection \cite{2023SoPh..298..114M,pianabook}. 
Then, we derive the coordinates of the flare location by identifying the position of the brightest area in the map, i.e., the location at which the X-ray emission is most intense. 
Note that the STIX image in Figure \ref{fig:bp} shows a background pattern around the flare location which does not represent a real flaring emission.
This pattern is a ringing effect due to the limited number of Fourier components available for image reconstruction and to the adopted image reconstruction method.

The resulting dataset is divided into training, validation, and test sets as follows. 
Samples constructed from files recorded between 2021 March 19 and 2022 November 15 (18K data points) are used for training, samples constructed from files recorded between 2022 November 15 and 2023 April 4 (13K data points) are used for validation, and the remaining ones (11K data points) are used for test. 
We did not use a random split of the dataset to avoid data leakage, where data points from the same flare would end up in both training and test sets. This would result in metrics overestimating the MLP performance.

\subsection{MLP implementation and training}\label{sec:MLP}

We implemented our MLP using the Python Tensorflow library \cite{tensorflow2015-whitepaper}.
Two normalization techniques appropriate for the requirements of the input and expected output data are applied prior to training the network. Input arrays are rescaled by their maximum value.
Normalizing the input values is permissible, as the intensity of the flare is not relevant when determining its location.
Similarly, the labels (i.e., the $x$ and $y$ coordinates of the flare locations) are divided by a value $f$ larger than the maximum dimension of the STIX Field-of-View (FOV). 
In this case, we set $f = 4000''$.
MLP training has been performed with the Adam optimizer \cite{kingma2014adam} and using a Mean Squared Error (MSE) loss function.

\begin{figure*}[t]

\centering
\includegraphics[width=.98\textwidth]{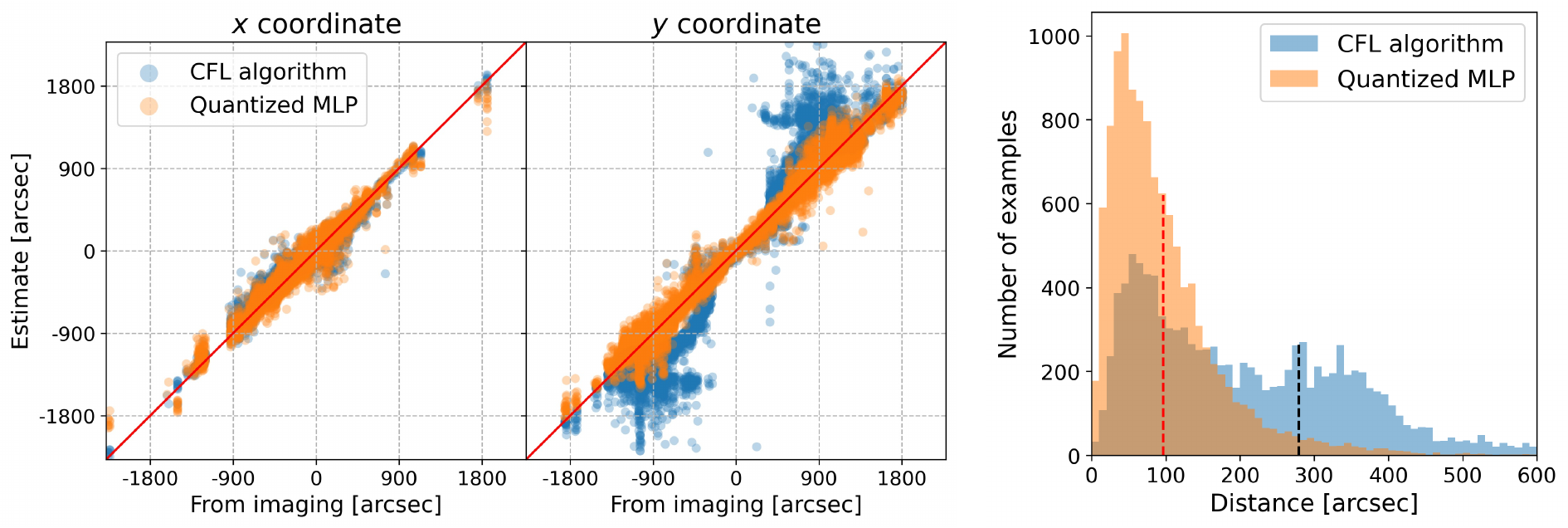}
\caption{Comparison of the quantized MLP and CFL algorithm performances on the test set examples.
Left and middle panels: scatter plots of the $x$ (left) and $y$ (middle) coordinates of the flare locations estimated by the MLP (orange) and by the CFL algorithm (blue) as a function of the coordinates obtained from the reconstructed STIX images. 
The identity line representing exact estimates is overlaid in red as a reference. 
Right panel: histograms of the distances between the flare locations estimated by the MLP (orange) and the CFL algorithm (blue) and the locations derived from STIX reconstructed images.
The mean values of the quantized MLP distances and of the CFL algorithm ones are plotted with a dashed red and black line, respectively.
}
\label{fig:2}

\end{figure*}

\subsection{Post-training quantization}
Since the STIX data processing unit does not support floating-point arithmetic, we apply a post-training quantization of inputs, weights, and outputs to ensure that all the network computations are performed in 16-bit fixed-point integer arithmetic at inference time.
As our specific hardware requirements are not satisfied by the available quantization algorithms, we use the custom approach described in the following.
We determined the range of all intermediate results by finding extremal values in a Mixed Integer Linear Programming representation of the network. A mathematical solver \cite{gurobi} proved that all numbers are strictly less than $2^5$, irrespective of the network's input. Thus, we use 5 bits to represent the integral part, and the remaining 11 bits to represent the fractional part of each number.
This way, our proposed model meets the computational requirements to be potentially run on-board the STIX instrument.

To convert the input values to fixed-point form, we use
\begin{equation} 
\mathbf{x}_{\mathrm{scaled}} = \left(\mathbf{x}_{\mathrm{input}} \cdot 2^{11} \right) ~//~  m ~,
\end{equation}
where $m$ is the maximum of the 9 input values, and $//$ denotes integer division (applied component-wise where needed). 
At the end of these operations, the input values are integers between 0 and $2^{11}$.

During the forward pass of the network, 16-bit input values are multiplied by 16-bit weights. The results are integers between 0 and $2^{32}$, which must be converted back to the 16-bit fixed-point representation by right shifting $11$ bits. Finally, we add the bias term and apply the ReLU activation function.
In summary, the output vector of a network layer is 
\begin{equation}
\mathbf{o} = \mathrm{ReLU}\left(\left(\mathbf{W} \cdot \mathbf{x} + 2^{10} \vec{\mathbf{1}} \right) \gg 11 + \mathbf{b}\right) ~,
\end{equation}
where $\mathbf{W}$ is the weight matrix, $\vec{\mathbf{1}}$ an all-one vector, $\mathbf{b}$ is the bias array, and $\gg$ denotes the component-wise right shift operator.
We note that the same operations are performed in the network's output layer, sans ReLU activation.
Finally, since inputs, weights, and biases of the network are scaled by $2^{11}$, and since ReLU is a positively homogeneous function of degree one (i.e., $\mathrm{ReLU}(\alpha \cdot \mathbf{x}) = \alpha \mathrm{ReLU}(\mathbf{x})$ for all $\alpha > 0$), the output of the network is also scaled by $2^{11}$. Thus the coordinates $\mathbf{c}$ computed by the network must be scaled back, also taking into account the fact that the output of the network during training has been divided by a value $f$ (see \cref{sec:MLP}). The array containing the estimated flare location coordinates is then
\begin{equation}
\mathbf{c}_{\mathrm{final}} = \left(\mathbf{c} \cdot f + 2^{10} \vec{\mathbf{1}} \right) \gg 11 ~,
\end{equation}
where $\mathbf{c}$ is the two-dimensional output of the network.

\subsection{CFL algorithm}\label{sec:CFL_algo}
During flaring events, the X-ray radiation is blocked by the tungsten ``H-shaped'' front grid of the CFL sub-collimator, and a shadow is cast on the corresponding detector (see Figure \ref{fig:1}).
The current on-board implementation of the CFL algorithm relies on a lookup table containing the values of the area of the
eight large CFL pixels that would be illuminated by
a point source located at each of 65 × 65 points on the solar
disk, uniformly separated by a distance of $2'$.
The CFL algorithm normalizes the eight large pixel measurements by an estimate of the total flux (i.e., total number of counts per unit of area) derived from the imaging sub-collimators, thus obtaining the fraction of illuminated area for each pixel.
Then, the method compares the observed illuminated areas with the entries of the lookup table, and returns the location corresponding to the maximum correlation.

In this proof of concept study, we utilize an on-ground Interactive Data Language (IDL) implementation of the CFL algorithm, and we apply it to our dataset (see Section \ref{sec:dataset}).
Differently from the on-board implementation, this version applies a final interpolation step between the coordinates corresponding to the largest correlation with the observations\footnote{The interpolation step is not performed in the on-board implementation of the CFL algorithm to reduce its computational cost.}.
Therefore, the on-ground implementation of the CFL algorithm can improve the $2'$ resolution of the lookup table.

\subsection{Results on test set} 

\cref{fig:2} compares the flare location coordinates obtained by means of the quantized neural network with those obtained by means of the CFL algorithm.
The left and middle panels show the scatter plots of the $x$ and $y$ coordinates of the flare locations obtained with the two methods as a function of the coordinates retrieved from the reconstructed STIX images.
The locations derived from imaging are the reference values; therefore, exact estimates would result in scatter plots along the identity diagonal, which is plotted in red as a reference.
The MLP predictions (orange) of the $x$ coordinate show an agreement with the locations derived from imaging which is comparable with that of the CFL algorithm (blue).
However, regarding the $y$ coordinate (middle panel), the MLP is substantially more accurate than the CFL algorithm.
This is particularly true for samples corresponding to flare locations $400'' < |y| < 1200''$ . 
In these locations one of the two rows of the CFL pixels is completely covered by the shadow of the ``H-shaped'' front grid (see the right panel of \cref{fig:1}).
To determine the amount of illuminated area in the large pixels, the CFL algorithm normalizes the measurements provided by the CFL pixels using an independent estimate of the total flux derived from the imaging sub-collimators (see Section \ref{sec:CFL_algo}).
It is likely that the cross-calibration between CFL measurements and those provided by the imaging sub-collimators is not yet optimal, which would explain the bias in the algorithm's estimate of the flare location. 
 
In the right panel of \cref{fig:2} we plot the histogram of the distances between the flare locations estimated by the methods and those derived from the reconstructed images.
The mean distance of the CFL algorithm predictions is $280''$, while the mean distance of the MLP predictions is $97''$, a marked improvement in accuracy.

\section{Discussion}\label{sec:3}

In this work, we develop a machine learning method for estimating the $(x,y)$ coordinates of flare locations on the solar disk from STIX CFL data.
Our model, coupled with a post-training quantization technique, is more accurate than the algorithm currently adopted on-board STIX.
Further, the CFL algorithm is based on a lookup table, which requires $65 \times 65 \times 8 \simeq \text{34K}$ parameters to be stored in memory, while our proposed MLP utilizes only 21K parameters (i.e., the total number of weights and biases).

The post-training quantization technique specifically developed for the adopted model architecture allows the MLP to use integer arithmetic at inference time, as required by the STIX data processing unit. 
The quantization technique does not affect the performance of the model. 
Indeed, the distance between the locations estimated by the quantized MLP and those estimated by the non-quantized model for the test set examples is, on average, $9''$, with a maximum distance of $24''$.

In this work we have considered an MLP with 3 layers and 100 neurons per layer. 
This choice of the model architecture was done based on some trial and error tests.
However, it is possible that even lighter models can be used for addressing the same task with similar accuracy.
Before considering using the trained neural network on-board STIX, we will more systematically search for the model architecture which provides the best trade-off between prediction accuracy of and low number of parameters by means of well-established techniques \cite{hinton2015distilling,white2023neural}.
In future work, we will also include flaring events outside the full-sensitivity STIX FOV in the training set.

Although more complex network architectures (e.g., convolutional neural networks) could be considered for this work, we decided to adopt an MLP model since the number of data points (9 elements) and number of output values (2 elements) is very small.
Given that the average error of the adopted MLP on the test set examples ($97''$) is already within the design accuracy of CFL sub-collimator (\(\sim\)$2'$), we do not expect more complex architectures to have substantially improved performances compared to the MLP.
Further, the MLP model was chosen as it could be easily implemented within the STIX flight software.
More complex model architectures would require a greater effort for the implementation on-board STIX, since the flight software can not leverage on standard Python libraries that are usually adopted for defining the models (e.g., Tensorflow \cite{tensorflow2015-whitepaper}).
Finally, we demonstrated that the MLP can be made compatible with the STIX hardware requirements by means of an appropriate post-training quantization technique.
The same quantization method can not be applied to every model.
Therefore, other models could require the implementation of tailored (and potentially complex) quantization methods.

The dataset and the code implemented for this study can be found at \url{https://github.com/paolomassa/STX_CFL_NN.git}.

\printbibliography

@article{krucker2020spectrometer,
  title={The Spectrometer/Telescope for Imaging X-rays (STIX)},
  author={Krucker, S{\"a}m and Hurford, Gordon J and Grimm, Oliver and K{\"o}gl, Stefan and Gr{\"o}belbauer, H-P and Etesi, L and Casadei, Diego and Csillaghy, Andr{\'e} and Benz, Arnold O and Arnold, Nicolas G and others},
  journal={\aap},
  volume={642},
  pages={A15},
  year={2020},
  publisher={EDP Sciences}
}

@ARTICLE{2020A&A...642A...1M,
       author = {{M{\"u}ller}, D. and {St. Cyr}, O.~C. and {Zouganelis}, I. and {Gilbert}, H.~R. and {Marsden}, R. and {Nieves-Chinchilla}, T. and {Antonucci}, E. and {Auch{\`e}re}, F. and {Berghmans}, D. and {Horbury}, T.~S. and {Howard}, R.~A. and {Krucker}, S. and {Maksimovic}, M. and {Owen}, C.~J. and {Rochus}, P. and {Rodriguez-Pacheco}, J. and {Romoli}, M. and {Solanki}, S.~K. and {Bruno}, R. and {Carlsson}, M. and {Fludra}, A. and {Harra}, L. and {Hassler}, D.~M. and {Livi}, S. and {Louarn}, P. and {Peter}, H. and {Sch{\"u}hle}, U. and {Teriaca}, L. and {del Toro Iniesta}, J.~C. and {Wimmer-Schweingruber}, R.~F. and {Marsch}, E. and {Velli}, M. and {De Groof}, A. and {Walsh}, A. and {Williams}, D.},
        title = "{The Solar Orbiter mission. Science overview}",
      journal = {\aap},
         year = 2020,
       volume = {642},
          eid = {A1},
        pages = {A1},
          doi = {10.1051/0004-6361/202038467},
       adsurl = {https://ui.adsabs.harvard.edu/abs/2020A&A...642A...1M}
}

@ARTICLE{2012NIMPA.695..288M,
       author = {{Meuris}, A. and {Hurford}, G. and {Bednarzik}, M. and {Limousin}, O. and {Gevin}, O. and {Le Mer}, I. and {Martignac}, J. and {Horeau}, B. and {Grimm}, O. and {Resanovic}, R. and {Krucker}, S. and {Orlea{\'n}ski}, P.},
        title = "{Caliste-SO X-ray micro-camera for the STIX instrument on-board Solar Orbiter space mission}",
      journal = {Nuclear Instruments and Methods in Physics Research A},
         year = 2012,
        month = dec,
       volume = {695},
        pages = {288-292},
          doi = {10.1016/j.nima.2011.11.016},
       adsurl = {https://ui.adsabs.harvard.edu/abs/2012NIMPA.695..288M}
}

@ARTICLE{2023SoPh..298..114M,
       author = {{Massa}, Paolo and {Hurford}, Gordon J. and {Volpara}, Anna and {Kuhar}, Matej and {Battaglia}, Andrea F. and {Xiao}, Hualin and {Casadei}, Diego and {Perracchione}, Emma and {Garbarino}, Sara and {Guastavino}, Sabrina and {Collier}, Hannah and {Dickson}, Ewan C.~M. and {Emslie}, A. Gordon and {Ryan}, Daniel F. and {Maloney}, Shane A. and {Schuller}, Frederic and {Warmuth}, Alexander and {Massone}, Anna Maria and {Benvenuto}, Federico and {Piana}, Michele and {Krucker}, S{\"a}m},
        title = "{The STIX Imaging Concept}",
      journal = {\solphys},
         year = 2023,
        month = oct,
       volume = {298},
       number = {10},
          eid = {114},
        pages = {114},
          doi = {10.1007/s11207-023-02205-7},
       adsurl = {https://ui.adsabs.harvard.edu/abs/2023SoPh..298..114M}
}

@book{pianabook,
author = {Piana, M. and Emslie, A.G. and Massone, A. M. and Dennis, B. R.},
title = {Hard X-ray Imaging of Solar Flares},
publisher = {Springer-Verlag},
year = {2022},
address = {Berlin}
}

@ARTICLE{2023A&A...673A.142X,
       author = {{Xiao}, Hualin and {Maloney}, Shane and {Krucker}, S{\"a}m and {Dickson}, Ewan and {Massa}, Paolo and {Lastufka}, Erica and {Francesco Battaglia}, Andrea and {Etesi}, L{\'a}szl{\'o} and {Hochmuth}, Nicky and {Schuller}, Fr{\'e}d{\'e}ric and {Ryan}, Daniel F. and {Limousin}, Olivier and {Collier}, Hannah and {Warmuth}, Alexander and {Piana}, Michele},
        title = "{The data center for the Spectrometer and Telescope for Imaging X-rays (STIX) on board Solar Orbiter}",
      journal = {\aap},
         year = 2023,
        month = may,
       volume = {673},
          eid = {A142},
        pages = {A142},
          doi = {10.1051/0004-6361/202346031},
archivePrefix = {arXiv},
       eprint = {2302.00497},
 primaryClass = {astro-ph.SR},
       adsurl = {https://ui.adsabs.harvard.edu/abs/2023A&A...673A.142X}
}

@software{xiao_2024_10973277,
  author       = {Xiao, Hualin},
  title        = {{pystixsim: an imaging simulator for the X-ray 
                   Imager/Telescope onboard Solar Orbiter STIX}},
  month        = apr,
  year         = 2024,
  publisher    = {Zenodo},
  version      = {v1.0},
  doi          = {10.5281/zenodo.10973277},
  url          = {https://doi.org/10.5281/zenodo.10973277}
}

@book{bishop2006pattern,
  title={Pattern recognition and machine learning},
  author={Bishop, Christopher M},
  year={2006},
  publisher={Springer}
}

@misc{tensorflow2015-whitepaper,
title={ {TensorFlow}: Large-Scale Machine Learning on Heterogeneous Systems},
url={https://www.tensorflow.org/},
note={Software available from tensorflow.org},
author={
    Mart\'{i}n~Abadi and
    Ashish~Agarwal and
    Paul~Barham and
    Eugene~Brevdo and
    Zhifeng~Chen and
    Craig~Citro and
    Greg~S.~Corrado and
    Andy~Davis and
    Jeffrey~Dean and
    Matthieu~Devin and
    Sanjay~Ghemawat and
    Ian~Goodfellow and
    Andrew~Harp and
    Geoffrey~Irving and
    Michael~Isard and
    Yangqing Jia and
    Rafal~Jozefowicz and
    Lukasz~Kaiser and
    Manjunath~Kudlur and
    Josh~Levenberg and
    Dandelion~Man\'{e} and
    Rajat~Monga and
    Sherry~Moore and
    Derek~Murray and
    Chris~Olah and
    Mike~Schuster and
    Jonathon~Shlens and
    Benoit~Steiner and
    Ilya~Sutskever and
    Kunal~Talwar and
    Paul~Tucker and
    Vincent~Vanhoucke and
    Vijay~Vasudevan and
    Fernanda~Vi\'{e}gas and
    Oriol~Vinyals and
    Pete~Warden and
    Martin~Wattenberg and
    Martin~Wicke and
    Yuan~Yu and
    Xiaoqiang~Zheng},
  year={2015},
}

@misc{kingma2014adam,
      title={Adam: A Method for Stochastic Optimization}, 
      author={Diederik P. Kingma and Jimmy Ba},
      year={2017},
      eprint={1412.6980},
      archivePrefix={arXiv},
      primaryClass={cs.LG},
      url={https://arxiv.org/abs/1412.6980}, 
}

@misc{gurobi,
  author = {{Gurobi Optimization, LLC}},
  title = {{Gurobi Optimizer Reference Manual}},
  year = 2023,
  url = "https://www.gurobi.com"
}

@misc{hinton2015distilling,
      title={Distilling the Knowledge in a Neural Network}, 
      author={Geoffrey Hinton and Oriol Vinyals and Jeff Dean},
      year={2015},
      eprint={1503.02531},
      archivePrefix={arXiv},
      primaryClass={stat.ML},
      url={https://arxiv.org/abs/1503.02531}, 
}

@misc{white2023neural,
      title={Neural Architecture Search: Insights from 1000 Papers}, 
      author={Colin White and Mahmoud Safari and Rhea Sukthanker and Binxin Ru and Thomas Elsken and Arber Zela and Debadeepta Dey and Frank Hutter},
      year={2023},
      eprint={2301.08727},
      archivePrefix={arXiv},
      primaryClass={cs.LG},
      url={https://arxiv.org/abs/2301.08727}, 
}
\addcontentsline{toc}{section}{References}

\section*{Acknowledgments}
Solar Orbiter is a space mission of international collaboration between ESA and NASA, operated by ESA. The STIX instrument is an international collaboration between Switzerland, Poland, France, Czech Republic, Germany, Austria, Ireland, and Italy.

PM, FPR, MSS, and AC are supported by the Swiss National Science Foundation in the framework of the project Robust Density Models for High Energy Physics and Solar Physics (rodem.ch), CRSII5\_193716.
HX is supported by the PRODEX experiment agreement No. 4000140488 for STIX. SK is partially supported by the Swiss National Science Foundation Grant 200021L\_189180.

\end{document}